\documentclass[showpacs,amsmath,amssymb,twocolumn,superscriptaddress,notitlepage,preprintnumbers,pra]{revtex4-1}

\usepackage[dvips]{graphicx}
\usepackage{amsmath,amssymb,amsthm,mathrsfs,amsfonts,dsfont}
 
\usepackage{dcolumn}
\usepackage{bm}

\usepackage{amsmath}
\usepackage{amssymb}
\usepackage{braket}
\usepackage{physics}
\usepackage{amsthm}
\usepackage{natbib}
\usepackage{mathtools}
\usepackage{diagbox}
\usepackage[utf8]{inputenc}
\usepackage[english]{babel}
\usepackage{scalerel}[2014/03/10]
\usepackage{stackengine}
\usepackage{algpseudocode}
\usepackage{algorithm}

\usepackage{tikz}
\usetikzlibrary{shapes.geometric, arrows}

\tikzstyle{process} = [rectangle, rounded corners,
minimum width=1cm, 
minimum height=1cm, 
text centered, 
draw=black, 
fill=orange!30]
\tikzstyle{arrow} = [thick,<-,>=stealth]

\usepackage{color}
\usepackage{orcidlink}

\usepackage{changes}

\usepackage{hyperref}
\hypersetup{colorlinks=true, linkcolor=blue, citecolor=blue, urlcolor=black }

\usepackage{qcircuit}

\graphicspath{{figures/}}

\newcommand{\ud}{\mathrm{d}}

\newcommand{\Id}{\ensuremath{\mathbf{I}}}

\begin{document}

\title{Critical behavior of Ising model by preparing thermal state on quantum computer}

\author{Xiaoyang Wang\orcidlink{0000-0002-2667-1879} }
\thanks{zzwxy@pku.edu.cn}
\affiliation{School of Physics, Peking University, Beijing 100871, China
}
\affiliation{Collaborative Innovation Center of Quantum Matter, Beijing 100871, China}
\affiliation{Center for High Energy Physics, Peking University, Beijing 100871, China}

\author{Xu Feng}
\affiliation{School of Physics, Peking University, Beijing 100871, China
}
\affiliation{Collaborative Innovation Center of Quantum Matter, Beijing 100871, China}
\affiliation{Center for High Energy Physics, Peking University, Beijing 100871, China}

\author{Tobias Hartung}
\affiliation{Northeastern University London, Devon House, St Katharine Docks, London, E1W 1LP, United Kingdom}

\author{Karl Jansen}
\affiliation{NIC, DESY Zeuthen, Platanenallee 6, 15738 Zeuthen, Germany}

\author{Paolo Stornati}
\affiliation{ICFO-Institut de Ciencies Fotoniques, The Barcelona Institute of Science and Technology, Av. Carl Friedrich Gauss 3, 08860 
Castelldefels (Barcelona), Spain}

\date{\today}
\begin{abstract}
   We simulate the critical behavior of the Ising model utilizing a thermal state prepared using quantum computing techniques. The preparation of the thermal state is based on the variational quantum imaginary time evolution (QITE) algorithm. The initial state of QITE is prepared as a classical product state, and we propose a systematic method to design the variational ansatz for QITE. We calculate the specific heat and susceptibility of the long-range interacting Ising model and observe indications of the Ising criticality on a small lattice size. We find the results derived by the quantum algorithm are well consistent with the ones from exact diagonalization, both in the neighborhood of the critical temperature and the low-temperature region.
\end{abstract}

\maketitle
\section{Introduction}

With the development of quantum devices and quantum algorithms, it is possible to solve problems on quantum computers that are hard for classical ones. Quantum computers have already been successfully implemented in many fields, including quantum chemistry, condensed matter physics and lattice field theory, see references~\cite{McArdle_20, Klco_2018, Klco_2020, Ciavarella_2021, Funcke_2022,Clemente_strategies,Banuls2020} as some examples. With the growing number of qubits and improved fidelities of quantum devices, more realistic physical models can be tackled, and the potential of quantum computers can be explored. As an example of application, in this article, we prepare the thermal state of the Ising model with a quantum algorithm at various temperatures, including points close to the critical temperature and the low-temperature region. To demonstrate the feasibility of our approach, we compare the quantum simulation results of the chosen physical quantities with the results from classical simulations.

Numerous algorithms have been proposed to enable a quantum computer to prepare a thermal state. These include the quantum thermal dynamic method, where the target system is coupled with a bath at equilibrium~\cite{Terhal_2000}, variational quantum algorithm based on the thermofield double state~\cite{Wu_2019,21npj_Sagastizabal}, as well as many quantum imaginary time evolution(QITE) algorithms such as the one utilizing Hubbard-Stratonovich transformation~\cite{Chowdhury_16}, QITE based on variational ansatz (QITE-ansatz)~\cite{McArdle_19}, QITE based on measurement (QITE-measure)~\cite{Motta_20} and QITE by performing coordinatewise optimization~\cite{Benedetti_21}. The scope of our research is to focus on the usage of noisy intermediate-scale quantum (NISQ) devices~\cite{Preskill_2018,Bharti22}. Given the presence of quantum noise, it is necessary to minimize the depth of the quantum circuits. We utilize the QITE-ansatz algorithm to generate thermal states in our research, as it has a relatively shallower circuit depth in comparison to other algorithms mentioned previously. In QITE-ansatz algorithm, the imaginary time evolution is carried out on a prior parameterized quantum circuit, and the parameters are evolved variationally. Thus, the parameterized quantum circuit is usually called variational ansatz. The variational ansatz is designed for ground state preparation in most references utilizing QITE-ansatz, such as~\cite{McArdle_19,Liu2021,Zoufal_2021}. Here, for thermal state preparation, we propose to construct a variational ansatz converted from quantum circuits utilized in QITE-measure~\cite{Motta_20}. The circuit in QITE-measure can also carry out imaginary time evolution, but the circuit depth is quite large. The circuit depth can be much reduced by converting the circuit into a variational ansatz. For example, when simulating the Ising model, the quantum circuits in QITE-measure have $\sim 100$ layers, while the variational ansatz circuits used in this work have less than 10 layers.

In this article, we study the long-range interacting Ising model. Long-range interaction between spins is introduced naturally in trapped-ion spin systems~\cite{Islam_2013}, and its dynamics can be simulated utilizing quantum simulation algorithms. The long-range interaction also leads to interesting physics such as confinement~\cite{Liu_2019} and meson scattering~\cite{vovrosh2022dynamical}. Meanwhile, the long-range interaction leads to effective dimensions that impact the system's critical behavior. Here, we calculate the specific heat of the long-range interacting Ising model near the critical point and in the low-temperature region.

This article is organized as follows. In section~\ref{sec:Long-range-interacting-Ising-model}, we introduce the long-range interacting Ising model and the measurement method of relevant physical quantities on a quantum computer. In section~\ref{sec:QITE}, we discuss the process of thermal state preparation using QITE-ansatz algorithm in detail, especially the method of variational ansatz design. In section~\ref{sec:numerical-result}, we present the numerical results and discuss the observed indications of the criticality. Finally, in section~\ref{sec:discussion-and-outlook}, we summarize the techniques used in this article and discuss the possible extension for further works.

\section{Long-range interacting Ising model}\label{sec:Long-range-interacting-Ising-model}

We consider the $D=2$ dimensional Ising model on a square lattice $\Lambda$ with long-range interactions. The Hamiltonian reads
\begin{align}
    H = -\sum_{i > j\in \Lambda}\frac{J}{r_{ij}^{\alpha}}Z_i Z_j-h\sum_i Z_i,
    \label{eq:long-range-interacting-Hamiltonian}
\end{align}
where $Z_i$ is the Pauli-$Z$ operator on the $i$th spin. $J$ is the bare coupling strength, and $\alpha$ denotes the range of the interaction. $h$ denotes the strength of the longitudinal external field. The distance $r_{ij}$ is defined by the Manhattan distance under periodic boundary condition(PBC): Assuming the position of spin $i$ on the square lattice is represented by integer vector $\vec{r}^{i}=(r^i_1,\ldots,r^i_D)$ and the volume of the lattice is $|\Lambda|=N_1\times \ldots \times N_D$, then
\begin{align}
    r_{ij}=\sum_{d=1}^D \min(|r^i_d-r^j_d|,N_d-|r^i_d-r^j_d|).
\end{align}
This Hamiltonian is a generalization of the interaction part of the Hamiltonian introduced in reference \cite{Liu_2019}. It reduces to the original nearest-neighbor Ising model~(NNIM) in the limit $\alpha\rightarrow \infty$. Because in the limit $\alpha\rightarrow \infty$, all the long-range couplings $J/r_{ij}^{\alpha}$ with $r_{ij}>1$ vanish, except for nearest-neighbor ones with $r_{ij}=1$.

The state of the Ising system at a finite temperature is described by the density operator. Its equilibrium state is the Gibbs state of which the density operator reads
\begin{align}
    \rho=\frac{1}{Z_{\beta}}e^{-\beta H},\quad Z_{\beta}\equiv \tr(e^{-\beta  H}).
    \label{eq:gibbs-state}
\end{align}
Here $\beta$ is the inverse temperature $\beta\equiv 1/(k_B T)$ and we define $K\equiv J\beta$ for later convenience. For an arbitrary observable $O$, its expectation value of the thermal state is given by
\begin{align}
    \langle O\rangle\equiv \tr(\rho O).
    \label{eq:thermal-expectation}
\end{align}
This article targets the case where the expectation values are evaluated for different $K$ and a zero external field $h=0$.

Now we exhibit observables to compute the Ising model's specific heat and susceptibility. Analyzing these measures allows us to examine the critical behavior of the Ising model. The specific heat is defined by the changing rate of the internal energy in a unit volume when varying the temperature $T$. It can be evaluated by the energy-fluctuation relation:
\begin{align}
    C_v\equiv\frac{1}{|\Lambda|}\frac{\partial \langle H\rangle}{\partial T}=\frac{1}{|\Lambda| T^2}\left[ \langle H^2\rangle- \langle H\rangle^2\right],
    \label{eq:energy-fluctuation-relation}
\end{align}
where the last expression can be derived by taking the Gibbs state Eq.~(\ref{eq:gibbs-state}) to evaluate the expectation values.

Similarly, the susceptibility is defined by the changing rate of the magnetization in a unit volume with respect to the external field strength $h$ (evaluated at $h=0$). The total magnetization is given by 
\begin{align}
    \langle M\rangle\equiv \langle Z_{tot}\rangle,
\end{align}
where $Z_{tot}\equiv\sum_i Z_i$, i.e., the sum of all the spins in the lattice. Then the susceptibility can be evaluated according to the susceptibility-fluctuation relation
\begin{align}
    \chi\equiv\frac{1}{|\Lambda|}\frac{\partial\langle M\rangle}{\partial h}\Big|_{h=0}=\frac{1}{|\Lambda|T}\left[ \langle Z_{tot}^2\rangle- \langle Z_{tot}\rangle^2\right].
    \label{eq:susceptibility-fluctuation-relation}
\end{align}

In summary, evaluating the specific heat and susceptibility is equivalent to calculating the expectation values of the corresponding operators. The operators to be measured include
\begin{align}
    H^2,H , Z_{tot}^2,Z_{tot}
    \label{eq:concerned-observable}
\end{align}
which can all be reduced to linear combinations of Pauli operators. To evaluate the expectation values of the above operators on quantum computers, we can generate the thermal state utilizing a quantum algorithm and then evaluate the expectation values of the Pauli operators. Notice that for the above operators, the elementary Pauli operators can be written as products of Pauli-$Z$ operators, so they commute and can be measured simultaneously on the quantum computer. Combined with the fact that the Hamiltonian in  Eq.~(\ref{eq:long-range-interacting-Hamiltonian}) consists of only Pauli-$Z$ operators, we can simplify the initial state to be evolved on quantum computer. It enables us to simulate the system on a larger lattice. For general models, such as the Ising model with a transversal field, the simplification does not hold. More details can be found in section~(\ref{sec:cps}).

\section{Thermal state preparation with quantum imaginary time evolution}\label{sec:QITE}
 
One can use the quantum imaginary time evolution(QITE) algorithm to prepare a thermal state~\footnote{We have also tried non-QITE-based variational quantum algorithm proposed in reference~\cite{Wu_2019,21npj_Sagastizabal}. We find its accuracy is not as good as the QITE-ansatz algorithm when the simulation is performed on a noiseless statevector quantum simulator}, as demonstrated in previous studies~\cite{McArdle_19,Motta_20}. This section provides an explanation of the QITE-ansatz algorithm. QITE-ansatz algorithm is designed to evolve an $N_q$-qubit quantum state $\ket{\psi(0)}$ to 
\begin{align}
    \ket{\psi(\tau)}=\frac{e^{-\tau H}\ket{\psi(0)}}{\sqrt{\bra{\psi(0)}e^{-2\tau H}\ket{\psi(0)}}},
    \label{eq:psi_QITE}
\end{align}
where $\tau$ is a real number denoting imaginary time. The denominator is a normalization factor to guarantee the evolution's unitarity. Assuming we have the quantum circuit to carry out the unitary evolution, then by choosing the initial state to be the maximally mixed state~(defined as the density operator) $\ket{\psi(0)}\bra{\psi(0)}=\Id/\mathbf{d}$~\footnote{Here we abuse bra-ket notation for the mixed state. It will be explained in detail in the subsection~(\ref{sec:initial-state-preparation})}~($\Id$ is the identity operator of the $\mathbf{d}\equiv 2^{N_q}$ dimensional Hilbert space), one finds the final state is the thermal state with inverse temperature $\beta=2\tau$
\begin{align}
    \ket{\psi(\tau)}\bra{\psi(\tau)}=  \frac{1}{Z_{2\tau}}e^{-2\tau H},\quad Z_{2\tau} \equiv \tr(e^{-2\tau  H}).
    \label{eq:QITE-thermal-state}
\end{align}

The QITE-ansatz algorithm was proposed in references \cite{McArdle_19, Yuan_2019}. This technique is originally used to project out the ground state of the Hamiltonian according to Eq.~(\ref{eq:psi_QITE}). It has been successfully implemented in the field of quantum chemistry, quantum field theory and machine learning, see e.g.~\cite{McArdle_20,Liu2021,Zoufal_2021}.

Following \cite{Yuan_2019}, we first review the QITE-ansatz algorithm within the density operator formalism. The density operator of Eq.~(\ref{eq:psi_QITE}) reads
\begin{align}
    \rho(\tau) = \frac{e^{-\tau H}\ket{\psi(0)}\bra{\psi(0)}e^{-\tau H}}{\bra{\psi(0)}e^{-2\tau H}\ket{\psi(0)}}.
\end{align}
The mathematical description of a quantum state with the density operator is equivalent to that with the pure state. In particular, the expectation values of any observable $O$ coincide
\begin{align}
\tr(\rho(\tau)O)=\bra{\psi(\tau)}O\ket{\psi(\tau)}.
\end{align}
The imaginary time evolution of the density operator follows the von-Neumann equation~\cite{Yuan_2019}
\begin{align}
    \frac{\ud \rho(\tau)}{\ud \tau}=\mathcal{L}[\rho(\tau)],
    \label{eq:von-Neumann equation}
\end{align}
where $\mathcal{L}$ is the Liouville operator defined by $\mathcal{L}(\rho)=-\{H,\rho\}+2\tr(\rho H)\rho$ with anti-commutator $\{H,\rho\}=H\rho+\rho H$. As the Hilbert space of the whole $N_q$ qubits is hard to be explored by a quantum circuit, we utilize a density operator $\hat{\rho}(\tau)=\ket{\phi(\tau)}\bra{\phi(\tau)}$ to approximate the target density $\rho(\tau)$. The approximation $\hat{\rho}(\tau)$ satisfies the following requirements: (1) It has the same initial state $\hat{\rho}(0)=\rho(0)=\ket{\psi(0)}\bra{\psi(0)}$. (2) The evolution of $\hat{\rho}(\tau)$ approximately satisfies the von-Neumann equation $\ud \hat{\rho}(\tau)/\ud \tau-\mathcal{L}[\hat{\rho}(\tau)]=0$.

The approximation $\hat{\rho}(\tau)$ is generated with a variational ansatz $\ket{\phi(\vec{\theta}(\tau))}=U(\vec{\theta}(\tau))\ket{\psi(0)}$, where $\vec{\theta}$ is a real variational parameter vector with $N$ components. $U(\vec{\theta})=U_N(\theta_N)\ldots U_1(\theta_1)$ is a series of parameterised unitary quantum gates. According to the first requirement mentioned above, $U(\vec{\theta}(0))$ should be the identity operator $\Id$. With the variational ansatz, the evolution of the quantum state is converted to the evolution of the variational parameters $\vec{\theta}$. However, as the variational ansatz cannot explore the whole Hilbert space, $\ket{\phi(\vec{\theta}(\tau))}$ can not fulfill the von-Neumann equation exactly. Instead, we demand that the von-Neumann equation is fulfilled sufficiently well according to the second requirement. The violation of the von-Neumann equation is measured by the McLachlan distance $L^2$, which is defined by
\begin{align}
    L^2\equiv \left|\left| \frac{\ud \hat{\rho}(\tau)}{\ud \tau}- \mathcal{L}[\hat{\rho}(\tau)]\right|\right|^2,
\end{align}
where $||A||^2=\tr(A^{\dagger}A)$ represents Frobenius norm. According to the differential chain rule, we have 
\begin{align}
    L^2=  \left|\left| \sum_{\mu} \frac{\partial \hat{\rho}(\theta)}{\partial \theta_{\mu}}\dot{\theta}_{\mu}-\mathcal{L}(\hat{\rho}) \right|\right|^2.
\end{align}
So that the McLachlan distance is a quadratic function of the time derivatives of the variational parameters $\dot{\theta}_{\mu}\equiv \partial \theta_{\mu}/\partial \tau$. $L^2$ can be minimized with the variational principle, which leads to
\begin{align}
    \delta L^2=0\Rightarrow \frac{\partial L^2}{\partial \dot{\theta}_{\mu}}=\sum_{\nu} M_{\mu\nu}\dot{\theta}_{\nu}-V_{\mu}=0,
\end{align}
where 
\begin{equation}
\begin{aligned}
    M_{\mu\nu}&\equiv 2\Re\left[\frac{\partial \bra{\phi(\vec{\theta})}}{\partial \theta_{\mu}}\frac{\partial \ket{\phi(\vec{\theta})}}{\partial \theta_{\nu}}\right],\\
    V_{\mu}&\equiv -2\Re\left[\frac{\partial \bra{\phi(\vec{\theta})}}{\partial \theta_{\mu}} H\ket{\phi(\vec{\theta})}\right].
    \label{eq:M-V-calculation}
\end{aligned}
\end{equation}
Here $M$ is a $N\times N$ matrix while $V$ is a $N$ dimensional vector. Following \cite{McArdle_19,Funcke2021dimensional}, one can construct some specific quantum circuits to measure $M$ and $V$, which cost $\mathcal{O}(N^2)$ quantum device calls and one additional ancilla qubit. 

After deriving $M$ and $V$, we can construct the following linear equations
\begin{align}
    \sum_{\nu} M_{\mu\nu}\dot{\theta}_{\nu}=V_{\mu}.
\end{align}
Then one can solve for the time derivative of the variational parameters $\dot{\theta}_{\nu}|_{\tau=\tau_0}$ at a given imaginary time $\tau_0$, utilizing methods such as pseudo-inverse~\cite{McArdle_19}. The variational parameters at the next time slice $\tau_0+\delta \tau$ are given according to the Euler method
\begin{align}
    \vec{\theta}(\tau_0+\delta \tau) \simeq \vec{\theta}(\tau_0)+\dot{\vec{\theta}}\delta\tau,
    \label{eq:Euler-integration}
\end{align}
where $\dot{\theta}_{\nu}=\sum_{\mu}M^{-1}_{\nu\mu}V_{\mu}$.

The computational complexity of the QITE-ansatz grows polynomially with the number of variational parameters $N$. In each time slice, the time complexity of solving linear equations grows polynomially with $N$, while the matrix $M$ and vector $V$ can also be evaluated using quantum computers within polynomial time. Thus as long as $N$ grows polynomially with the system size $N_q$, the time complexity of the QITE-ansatz grows polynomially with $N_q$ and can be extended to large-scale quantum systems. The following subsections will introduce how to prepare the maximally mixed state and choose an appropriate variational ansatz.

\subsection{Initial state preparation}\label{sec:initial-state-preparation}\label{sec:cps}
Here we introduce how to prepare the initial state as the maximally mixed state $\Id/\mathbf{d}$. Quantum circuits are suitable for generating pure states. We need some strategies to generate mixed states utilizing pure states. As discussed in \cite{White2009}, there are two strategies: ancilla pair state~(APS) and classical product state~(CPS). Both strategies can be used to prepare maximally mixed state $\Id/\mathbf{d}$. However, preparing $\Id/\mathbf{d}$ with APS doubles the number of qubits to $2N_q$~\cite{Zoufal_2021}. It also introduces some complexities in variational ansatz design to evolve the pair state. 

Instead, we can prepare the maximally mixed state via CPS, which reduces the required qubits to $N_q$. The maximally mixed state $\Id/\mathbf{d}$ describes that the probabilities of sampling every basis vector from a given orthogonal basis are the same, where each basis vector is a pure state. As the maximally mixed state is unitarily invariant $U(\Id/\mathbf{d})U^{-1}=\Id/\mathbf{d}$, the orthogonal basis can be chosen arbitrarily. To generate the thermal state, it is recommended in \cite{White2009} to use a basis formed by classical product states, such as $\{\ket{+},\ket{-}\}^{\otimes N_q}$, where $\{\cdot\}^{\otimes N_q}$ represents a set generated by the $N_q$ times tensor product of each element in $\{\cdot\}$. For example,
\begin{align}
    \{\ket{+},\ket{-}\}^{\otimes 2}=\{\ket{++},\ket{+-},\ket{-+},\ket{--}\}.
    \label{eq:X-basis-vector}
\end{align}
Here $\ket{+}$,$\ket{-}$ represent the eigenvectors of the Pauli-$X$ operator
\begin{align}
    X\ket{+}=\ket{+},\quad X\ket{-}=-\ket{-}.
    \label{eq:X-basis}
\end{align}

If we use the classical product state as the initial state, the thermal expectation value $\langle O\rangle$ can not be measured straightforwardly due to the normalization factor in Eq.~(\ref{eq:psi_QITE}). Assume that we take the orthogonal basis as $\{\ket{i}\}$. Evolving all basis vectors $\ket{i}$ for imaginary time $\tau$, one gets the expectation values of an observable $O$, which read
\begin{align}
    \bra{i(\tau)}O\ket{i(\tau)}=\frac{\bra{i}e^{-\tau H}Oe^{-\tau H}\ket{i}}{\bra{i}e^{-2\tau H}\ket{i}}.
    \label{eq:expectation_CPS}
\end{align}
Usually, the denominators would be different for different basis vectors $\ket{i}$. To derive the thermal expectation value $\langle O\rangle$ in Eq.~(\ref{eq:thermal-expectation}), we should multiply the above expectation values with coefficients $\{p_i\}$
\begin{align}
    \langle O\rangle =\sum_i p_i \bra{i(\tau)}O\ket{i(\tau)},
    \label{eq:CPS-expectation}
\end{align}
where $p_i$ is defined by
\begin{align}
    p_i\equiv\frac{\bra{i}e^{-2\tau H}\ket{i}}{Z_{2\tau}}.
\end{align}
Here $\{p_i\}$ can be treated as a probability distribution, as they are all positive and satisfy the normalization condition $\sum_i p_i=1$. To evaluate the thermal expectation value of the operator $O$, as mentioned in \cite{Motta_20}, we do not need to calculate all the $\{p_i\}$(which would be impossible to calculate, as the number of $p_i$ grows exponentially with the number of qubits). With the minimally entangled typical thermal state(METTS) algorithm proposed by Stoudenmire and White~\cite{Stoudenmire_2010}, one can sample $\{\ket{i}\}$ according to the distribution $\{p_i\}$. The thermal expectation value $\langle O\rangle$ is the average of the expectation of $O$ with the time-evolved sampled vectors. In conclusion, though imaginary time evolution with CPS as initial states requires the number of qubits equal to the system size, one has to evolve different initial states $\ket{i}$ to acquire statistics. On the other hand, imaginary time evolution with APS as an initial state doubles the number of qubits while evolving only one initial state.

However, the situation gets simplified when we consider the classical Ising model and the observables in Eq.~(\ref{eq:concerned-observable}), which consist of Pauli-$Z$ operators. The observables can be generally expressed as
\begin{align}
    O=\sum_m h_m\tilde{Z}_m.
    \label{eq:observable-Z-decomposition}
\end{align}
Here $\tilde{Z}_m$ represents the tensor product of $Z$ operators at some sites and identity operators at the others, such as $\tilde{Z}_m=Z_{N_q-1}\ldots I_1Z_0$. In Appendix~\ref{app:cps-expectation}, we prove that the thermal expectation value of $O$ can be calculated according to 

\begin{equation}
\begin{aligned}
     \langle O\rangle=\sum_m h_m \bra{\tilde{+}(\tau)}\tilde{Z}_m\ket{\tilde{+}(\tau)},
     \label{eq:simplified-estimation}
\end{aligned}
\end{equation}
where $\ket{\tilde{+}(\tau)}$ is imaginary time evolved state according to Eq.~(\ref{eq:psi_QITE}). The state is initialized as $\ket{\tilde{+}(0)}=\ket{\tilde{+}}$, where $\ket{\tilde{+}}\equiv\ket{+}^{\otimes N_q}$ is the $N_q$-fold tensor product of $\ket{+}$ in Eq.~(\ref{eq:X-basis}). Thus for the Ising model, we only need to calculate the imaginary time evolution with the initial state $\ket{\tilde{+}}$.

In this work, we use $\ket{\tilde{+}}$ as the initial state to present our results. For general models, such as the Ising model with a transversal field, the above simplification does not hold. We need to sample the classical product states using the METTS algorithm or utilize the ancilla pair state. 

\subsection{Variational ansatz design}
Choosing a proper variational ansatz is a cornerstone for the success of the QITE-ansatz algorithm~\cite{Bharti22}. In most literature on QITE-ansatz, the variational ansatz is designed to prepare the ground state of a Hamiltonian, and it is suitable to evolve some specific initial states, such as the unitary coupled cluster ansatz evolving Hartree-Fock states~\cite{McArdle_20}. Focusing on thermal state preparation and the initial state introduced in the previous section, we propose to construct a variational ansatz converted from quantum circuits utilized in the QITE-measure algorithm proposed by Motta et. al.~\cite{Motta_20}. 

 We briefly introduce how to construct the quantum circuits used in the QITE-measure algorithm. The goal of QITE-measure is also evolving an initial state $\ket{\psi(0)}$ according to Eq.~(\ref{eq:psi_QITE}). Consider evolving the state $\ket{\psi(\tau_0)}$ for a small time slice $\Delta\tau$ 
\begin{align}
    \ket{\psi(\tau_0+\Delta\tau))}=\frac{e^{-\Delta\tau H}\ket{\psi(\tau_0)}}{\sqrt{\bra{\psi(\tau_0)}e^{-2\Delta\tau H}\ket{\psi(\tau_0)}}}.
\end{align}
As this transformation is unitary, we can always find a Hermitian operator $\hat{A}(\tau_0)$ such that
\begin{align}
    \ket{\psi(\tau_0+\Delta\tau)}=e^{-i\Delta \tau \hat{A}(\tau_0)}\ket{\psi(\tau_0)},
    \label{eq:psi_measurement_QITE}
\end{align}
and $\hat{A}(\tau_0)$ can be expanded in a complete Pauli basis
\begin{align}
    \hat{A}(\tau_0)=\sum_{i_1\ldots i_{N_q}} a_{i_1\ldots i_{N_q}}^{(\tau_0)} \sigma_{i_1}\ldots \sigma_{i_{N_q}}\equiv \sum_{I} a_{I}^{(\tau_0)} \tilde{\sigma}_I,
    \label{eq:Hermitian-expansion}
\end{align}
where the expansion coefficients $ a_{i_1\ldots i_{N_q}}^{(\tau_0)}$ are real due to the Hermicity of $\hat{A}(\tau_0)$, and $\sigma_{i_j}=I,X,Y,Z$ corresponding to $i_j=0,1,2,3$ is the single-qubit Pauli operator on the site $j$, and we call the tensor product of the single-qubit Pauli operator, $\tilde{\sigma}_I$ as Pauli string. For this reason, the single-qubit Pauli operator is sometimes called Pauli letter~\cite{Oliver_22}. For each imaginary time $\tau_0$, one can calculate all the expansion coefficients $a_{I}^{(\tau_0)}$ by evaluating the expectation values of some observables with respect to the quantum state $\ket{\psi(\tau_0)}$. The observables are the composition of Pauli strings and the Hamiltonian (See more details in \cite{Motta_20}). Notice that the transformation in Eq.~(\ref{eq:psi_measurement_QITE}) can be approximated by 
\begin{align}
    e^{-i\Delta \tau \sum_{I} a_{I}^{(\tau_0)} \tilde{\sigma}_I}=\prod_I e^{-i\Delta \tau a_{I}^{(\tau_0)} \tilde{\sigma}_I}+\mathcal{O}(\Delta \tau^2),
\end{align}
where the product consists of several Pauli exponentials which have the form $e^{-i\theta \tilde{\sigma}_I}$, and the Pauli exponential can be realized with quantum gates in a standard way~\cite{Nielsen2000}. Thus, the whole quantum circuit used in the QITE-measure can be constructed using several Pauli exponentials for each time slice. In the last time slice, the circuit depth is proportional to the final imaginary time $\tau$.

Notice that if a system has $N_q$ qubits, the total number of Pauli strings on these qubits is $4^{N_q}$. Thus the number of Pauli exponentials required for evolving each time slice seems exponential as a function of system size according to Eq.~(\ref{eq:Hermitian-expansion}). However, the situation gets simplified when the Hamiltonian $H$ consists of some local interaction terms
\begin{align}
    H=\sum_m H_m,
\end{align}
where each $H_m$ acts on a local set of qubits, and the number of $H_m$ is polynomial as a function of system size. For example, $H_m\propto Z_i Z_j$ and the number of $H_m$ is $\mathcal{O}(N_q^2)$ in case of long-range interacting Ising model. Though the local terms $H_m$ may not commute, the imaginary time evolution $e^{-\Delta \tau H}$ can be decomposed by
\begin{align}
    e^{-\Delta \tau H}=\prod_m e^{-\Delta \tau H_m}+\mathcal{O}(\Delta \tau^2).
\end{align}
Then the previous steps in QITE-measure can be implemented for each $e^{-\Delta \tau H_m}$. As shown in \cite{Motta_20}, when the Hamiltonian consists of local terms and the correlation length of the system is finite, the expansion in Eq.~(\ref{eq:Hermitian-expansion}) for each $H_m$ can be implemented with Pauli strings on a support constantly larger than the support of $H_m$ (Support of a Pauli string is defined by the set of qubits on which the Pauli letters are not identity). The correlation length of a system is finite when its Hamiltonian is outside the critical region. Thus the support of the Pauli strings has no dependence on the system size, and the total number of Pauli exponentials $e^{-i\theta \tilde{\sigma}_I}$ is a polynomial function of the system size at least when the Hamiltonian is sufficiently far away from the critical point.

Compared with the QITE-ansatz, the precision of the QITE-measure is not limited by the variational ansatz. However, the circuit depth grows linearly with the evolution time $\tau$. Thus this algorithm would be very sensitive to coherent or incoherent noise in real quantum devices and can only be applied to small spin systems~\cite{Aydeniz_2020}. 

Quantum circuits constructed in QITE-measure can be naturally converted into a variational ansatz with the following steps: (1) using all the necessary Pauli exponentials at one time slice as one layer of the variational ansatz; (2) sequentially repeating the layer several times in the quantum circuit; (3) converting all the expansion coefficients $a_{I}^{(\tau_0)}$ into undetermined parameters, which are initially zero and to be evolved according to the QITE-ansatz algorithm. Times of repetition for one layer is called the depth of the variational ansatz, also called the number of layers. 

The behavior of this variational ansatz can be analyzed with the help of QITE-measure. Assuming we have the same quantum circuit layers for the variational ansatz in QITE-ansatz and the quantum circuits in QITE-measure. Because the states prepared in QITE-measure can all be explored by the variational ansatz, one can expect QITE-ansatz using this circuit to behave at least better than QITE-measure. The systematic error of the QITE-measure circuit is of the first-order Trotter type, i.e., $\mathrm{error}\sim \mathcal{O}(\Delta \tau)$~\cite{Motta_20}. By equalizing the longest circuit depth used in QITE-measure and the depth in variational ansatz, it can be deduced that in the worst case, the variational ansatz leads to an error of $\mathcal{O}(1/\mathrm{L})$, where $L$ is the number of layers.



\begin{figure*}
\begin{center}
\includegraphics[width=0.85\textwidth]{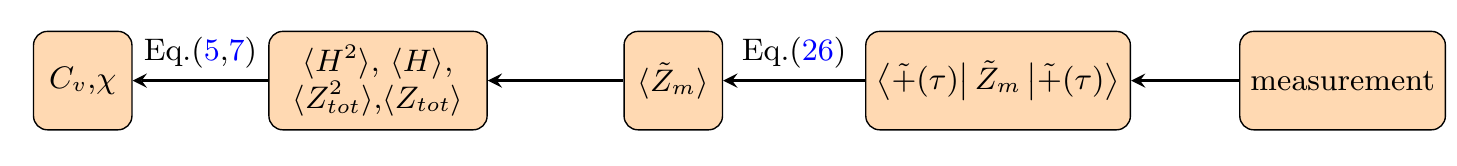}
\end{center}
    \caption{An illustration of the measurement process. The specific heat $C_v$ and susceptibility $\chi$ of Ising model can be calculated by the measurement in computational basis on quantum computers.}
    \label{fig:measurement-process}
\end{figure*}

In the numerical simulations, we find that the circuit depth required in QITE-ansatz is much smaller than that required in the QITE-measure. For example, in our numerical simulation of the Ising model, if the imaginary time of the final state is $\tau=0.5$, with step size $\Delta \tau=0.002$, QITE-measure requires the number of layers $\tau/\Delta \tau=250$. In contrast, to reach a sufficiently good precision using the variational ansatz, we find the number of layers required is at most $L=N_d$ for the 2-D nearest neighbor Ising model where $N_d$ is the side length of the Ising lattice system. More details on the number of circuit layers required are shown in Appendix~\ref{app:error-and-layer-estimation}.

The variational ansatz can be simplified due to some special structures of the Hamiltonian and the initial state. In the numerical simulations, we notice that some of the variational parameters are always zero during the whole evolution, which corresponds to the same set of Pauli strings over all the layers. We call the Pauli string in this set \textit{irrelevant}, and the other Pauli strings corresponding to non-zero variational parameters are \textit{relevant}. As the irrelevant Pauli exponentials are identity, they can be removed a priori when constructing the variational ansatz. These irrelevant Pauli strings can be identified according to the symmetry and some special structures of the Hamiltonian and the initial state. For example, if all the entries in the Hamiltonian and the initial state are real, then the corresponding unitary operator $e^{-i\Delta \tau \hat{A}}$ should also be real. Thus all Pauli strings with an even number of Pauli-$Y$ letters are irrelevant.

\begin{figure*}
    \centering
    \includegraphics[width = 0.9\textwidth]{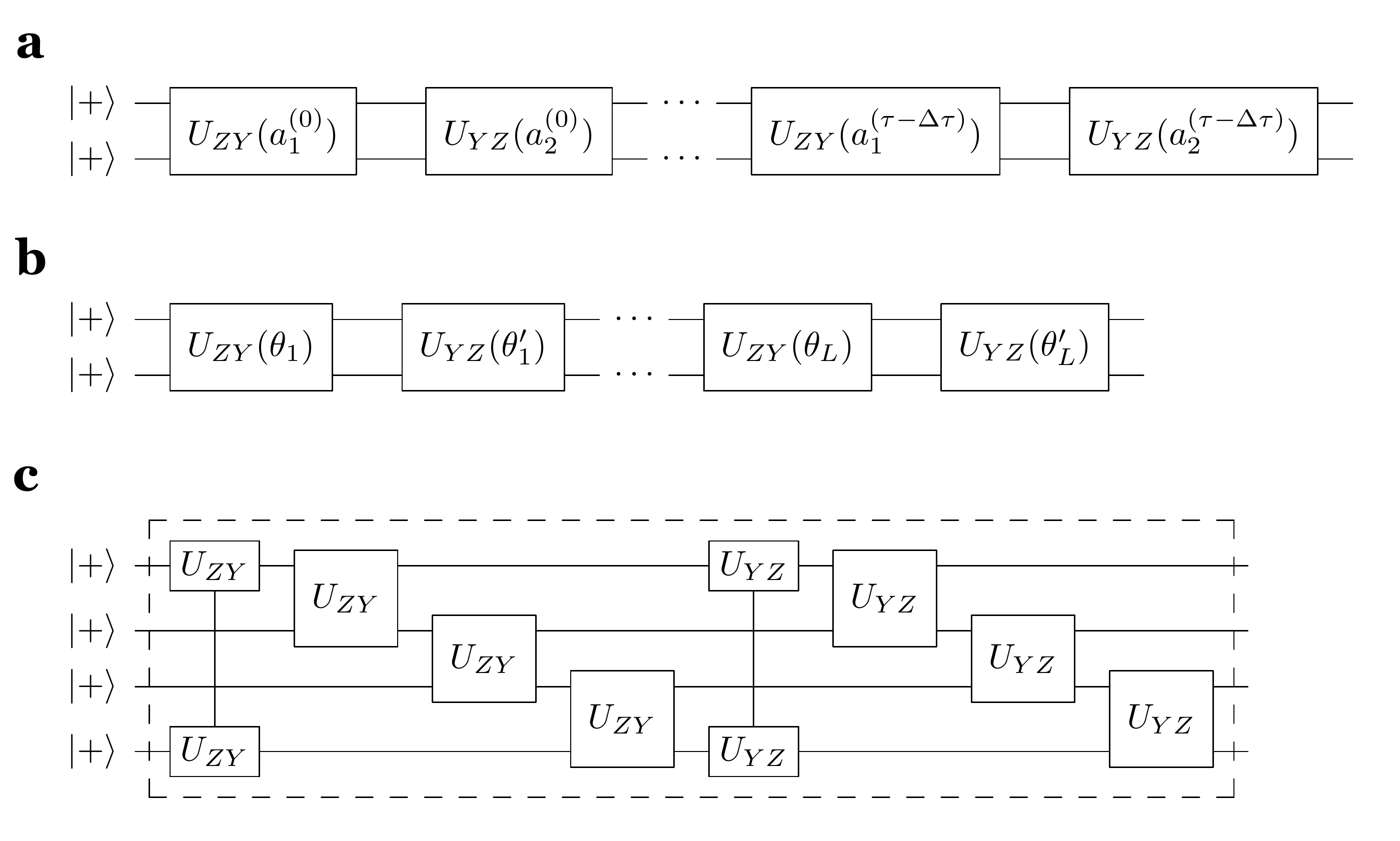}
    \caption{Example of circuits for the imaginary time evolution of Ising systems. The basic building blocks of the circuits are defined as $U_{ZY}(\theta)\equiv e^{-i\theta ZY}, U_{YZ}(\theta)\equiv e^{-i\theta YZ}$. (\textbf{a}) The quantum circuit in the QITE-measure algorithm to carry out the imaginary time evolution $e^{\tau ZZ}\ket{++}$. $\Delta \tau$ is the length of one time slice. (\textbf{b}) The variational ansatz converted from the QITE-measure circuit. $L$ is the number of circuit layers, $\theta_i,\theta_i', i\in[1,L]$ are free variational parameters. (\textbf{c}) The variational ansatz for nearest neighbor 1-D Ising chain under the periodic boundary condition. Each layer consists of one layer of ZY-Pauli exponentials and one layer of YZ-Pauli exponentials, as shown in the dashed box. The figure shows the case of one layer.}
    \label{fig:circuit}
\end{figure*}

We demonstrate the above construction of variational ansatz using an example of a two-qubit~($N_q=2$) Ising system. There are $4^2=16$ Pauli strings on the two-qubit system. Assume we have an initial state $\ket{++}$ and the system Hamiltonian $H=-Z_1Z_0$. Because all the entries in the Hamiltonian and the initial state are real, eliminating Pauli strings with an even number of Pauli-$Y$ letters leaves 6 Pauli strings: $I_1Y_0, X_1Y_0, Y_1I_0, Y_1X_0, Z_1Y_0, Y_1Z_0$. Evolving one layer with these 6 Pauli strings using QITE-ansatz, we further find 4 Pauli strings are irrelevant. It leaves only two relevant Pauli strings for the imaginary time evolution
\begin{align}
    Z_1Y_0,Y_1Z_0.
\end{align}
One can verify that
\begin{equation}
\begin{aligned}
    e^{-\Delta \tau H}\ket{++}&=e^{\Delta \tau Z_1 Z_0}\ket{++}\\
    &\propto e^{-ia_1(0) Z_1 Y_0}e^{-ia_2(0) Y_1 Z_0}\ket{++},
    \label{eq:two-qubit-QITE}
\end{aligned}
\end{equation}
with expansion coefficients
\begin{align}
    a_1^{(0)}=a_2^{(0)}=\frac{1}{2}\tan^{-1}(\tanh \Delta \tau).
\end{align}
In the QITE-measure algorithm, to evolve the initial state to an arbitrary time $\tau$, the quantum circuit is shown in figure~\ref{fig:circuit}\textbf{a}. It has $\tau/\Delta \tau$ layers. The variational ansatz with $L$ layers for the two-qubit Ising system is constructed as shown in figure~\ref{fig:circuit}\textbf{b}. In this circuit, $\{\theta_1,\theta_1'\ldots \theta_L,\theta_L'\}$ are all variational parameters, taking zero as initial values, and to be evolved according to the QITE-ansatz algorithm. 
\begin{algorithm}[H]
\caption{QITE-ansatz algorithm for Ising model}\label{alg:qite-ansatz}
\begin{algorithmic}
 \State  Construct ansatz $\ket{\phi(\vec{\theta}(\tau))}$ for the Ising Hamiltonian $H$.
 \State  Initial inverse temperature $K = 2\tau \gets 0$. 
 \State  Rotation angles $\vec{\theta}(0)\gets \vec{0}$. 
 \State  Initial state $\ket{\phi(\vec{0})}\gets \ket{\tilde{+}}$
 \While{$\tau\leq \tau_{\max}$}
  \State Calculate specific heat and susceptibility by the measurement process in Figure~\ref{fig:measurement-process}.
  \State Calculate matrix $M$ and vector $V$ in Eq.~(\ref{eq:M-V-calculation})
   \State Update rotation angles: $\vec{\theta}(\tau+\delta\tau) \gets \vec{\theta}(\tau)+\delta\tau \cdot M^{-1} V.$
   \State $\tau \gets \tau+\delta \tau$
 \EndWhile 
\end{algorithmic}
\end{algorithm}

\section{Numerical results}\label{sec:numerical-result}

\begin{figure*}
    \centering
    \includegraphics[width = 0.40\textwidth]{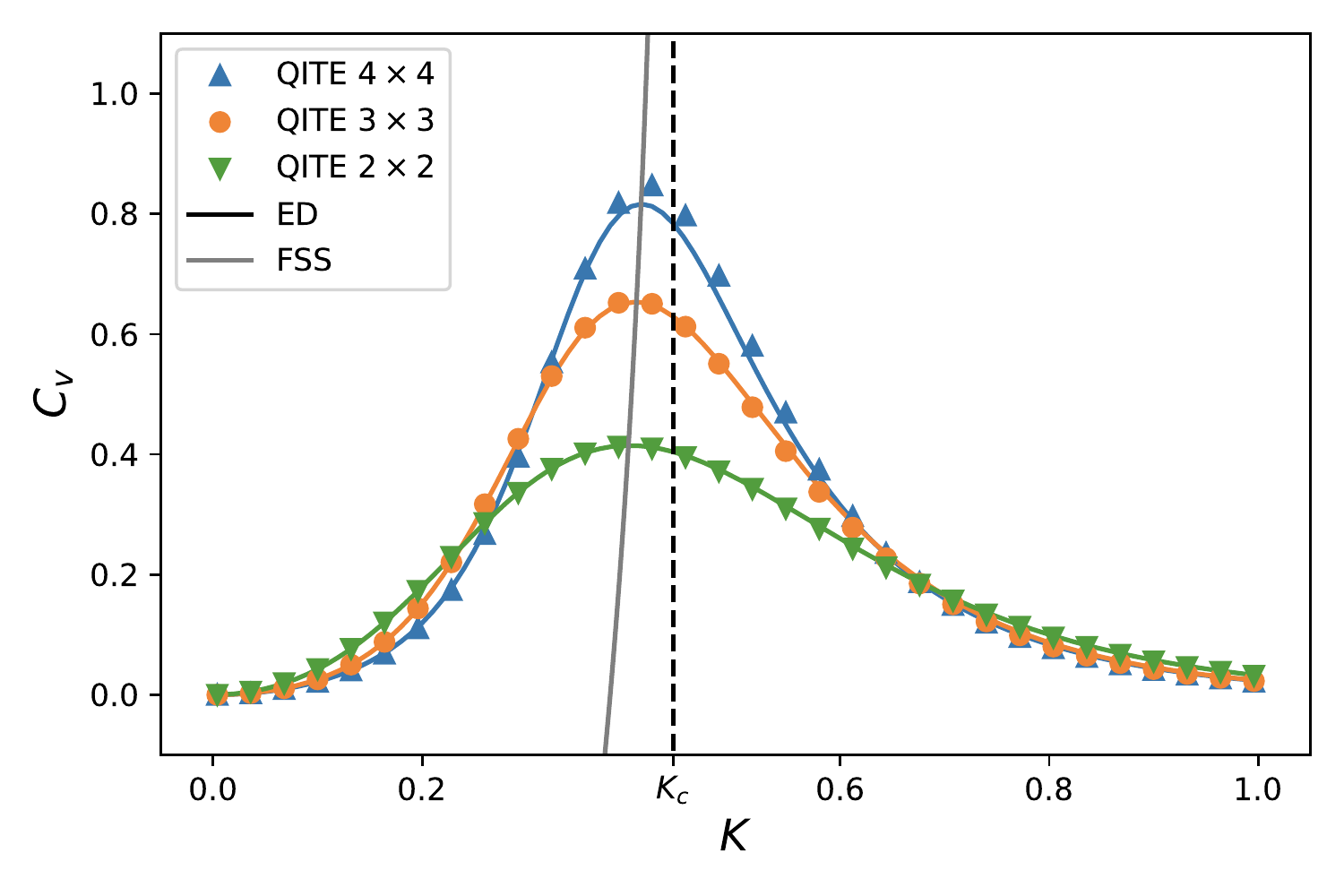}
    \includegraphics[width = 0.40\textwidth]{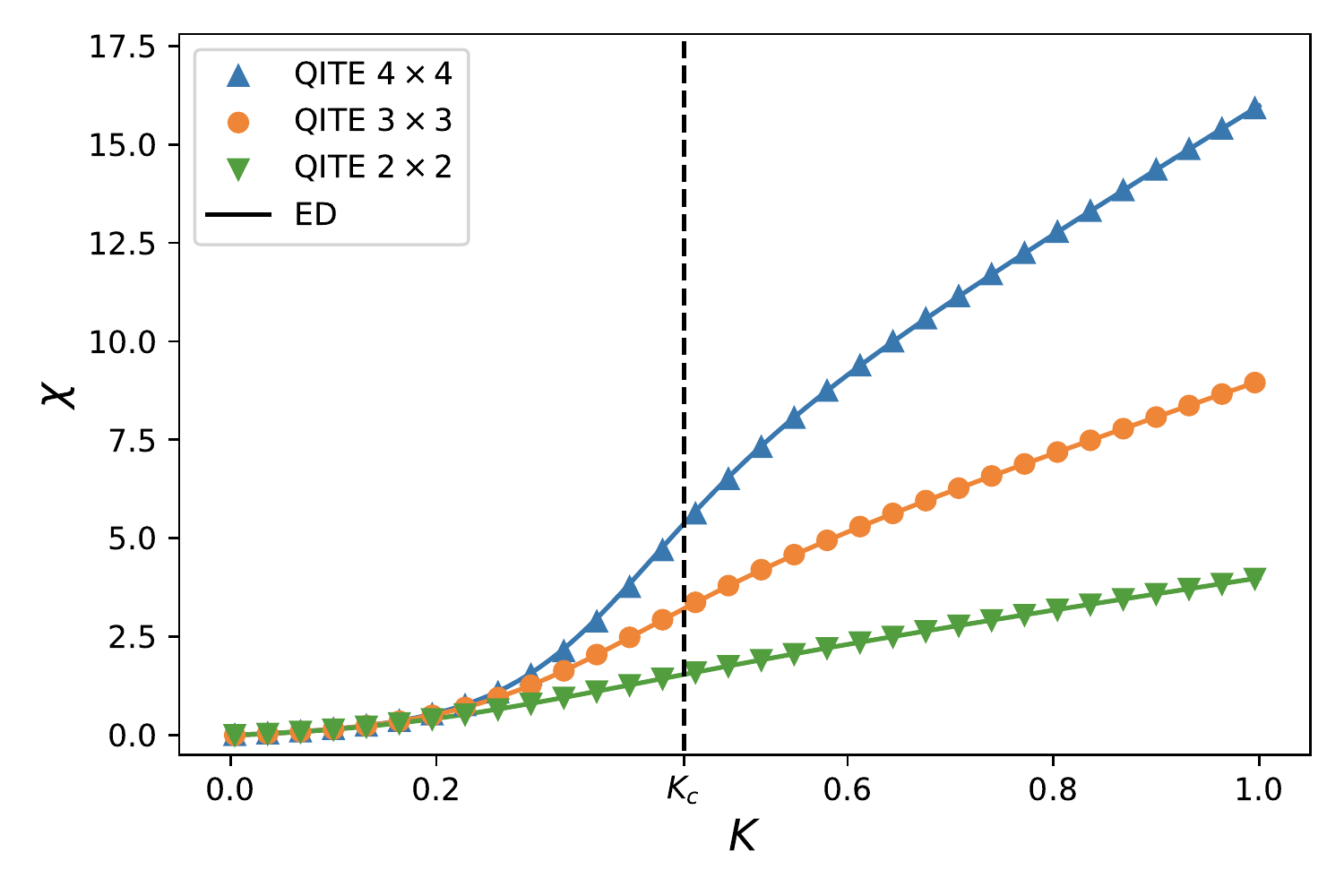}
    \includegraphics[width = 0.40\textwidth]{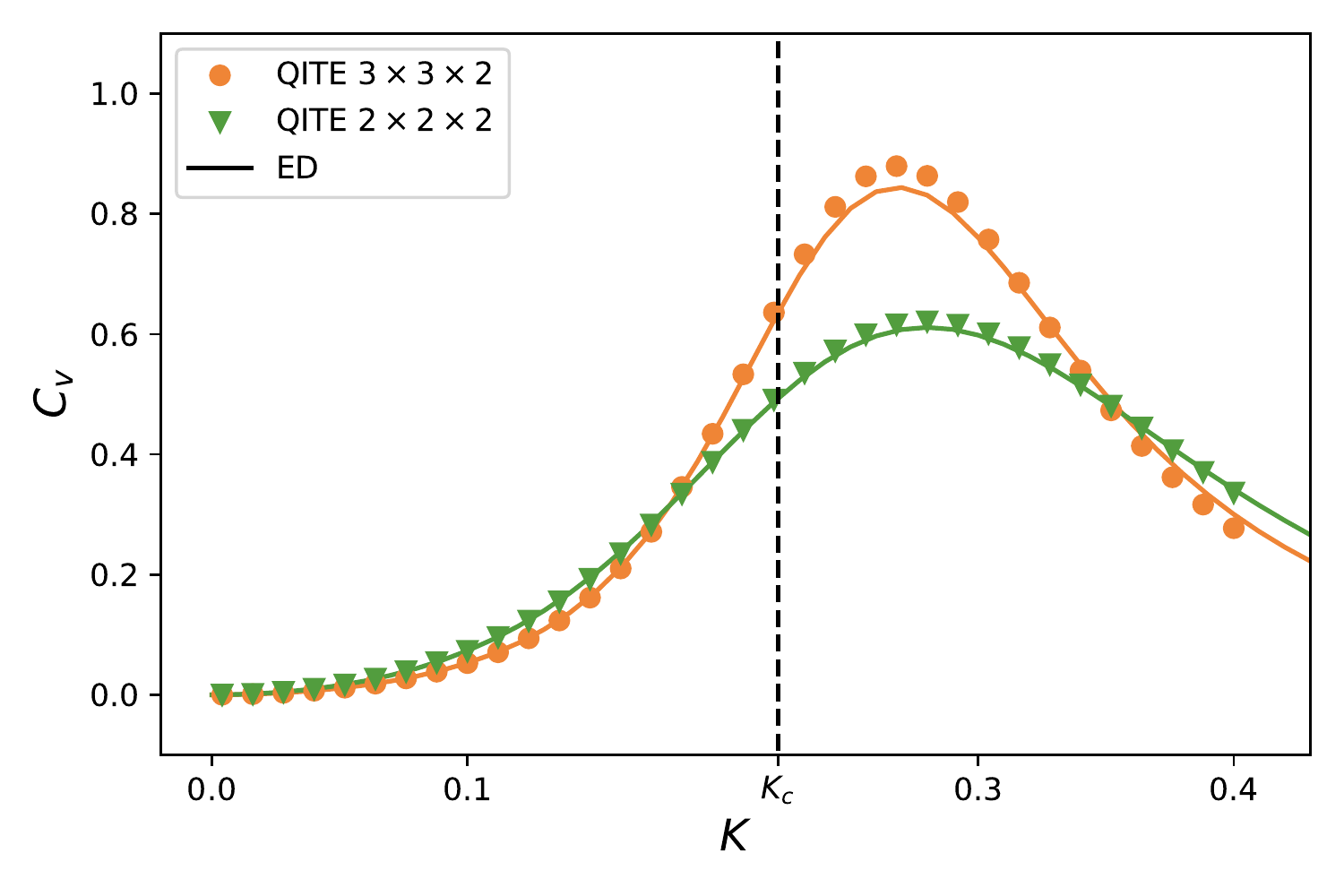}
    \includegraphics[width = 0.40\textwidth]{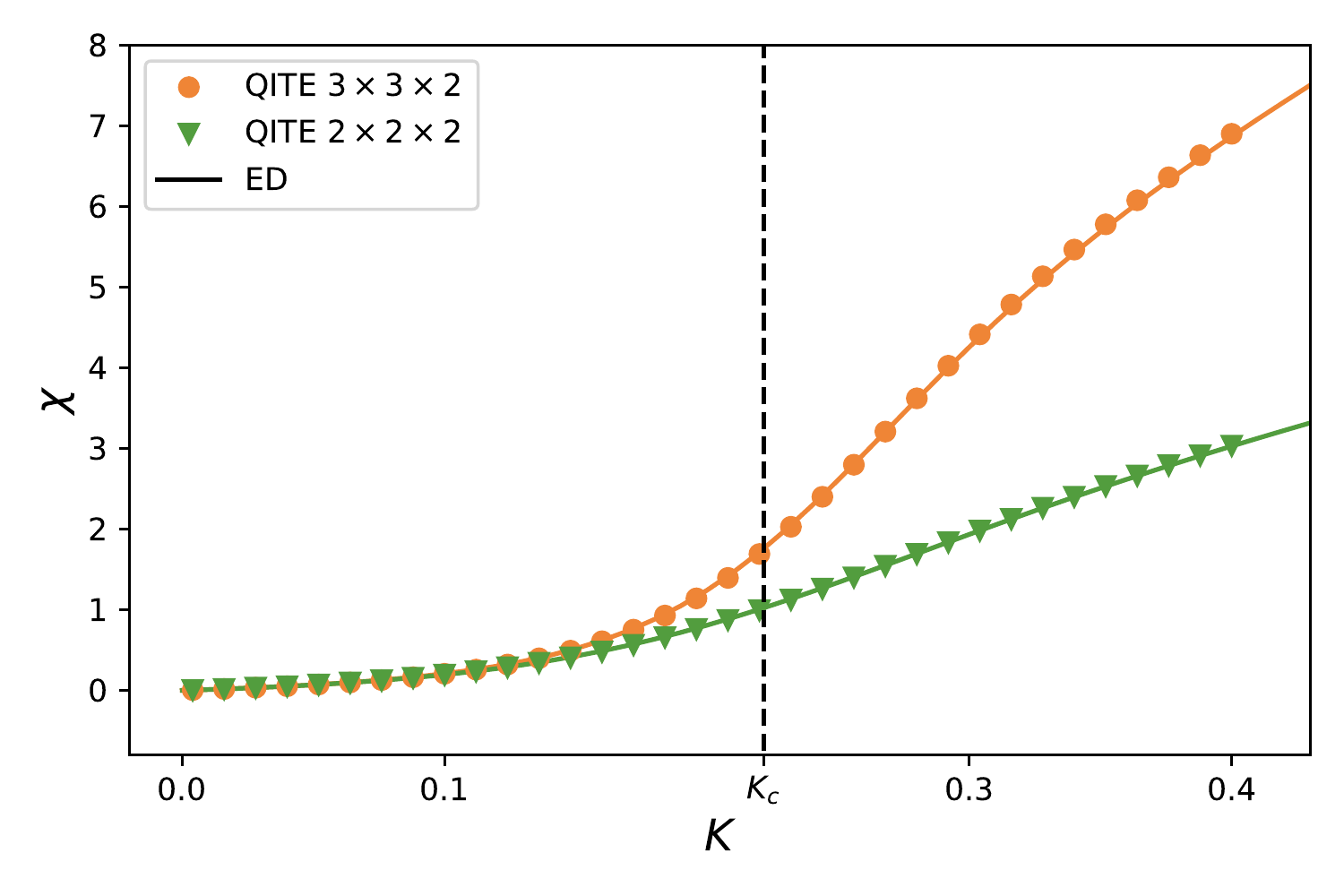}
    \caption{Specific heat(left column) and susceptibility(right column) as a function of $K$ in 2-D(upper row) and 3-D(lower row) nearest-neighbor Ising model($\alpha=\infty$). ED represents results from exact diagonalization. We see that the results from the noiseless quantum simulation are close to exact diagonalization, especially in the region far from the critical point. The black dashed lines in the four panels are the exact critical temperatures $K_c$ of the corresponding dimension in the infinite volume limit. The solid grey line in the upper left panel shows the peak movement as the system size enlarged, which is fitted inspired by finite size scaling~(FSS). As the lattice size increases, the peaks of the specific heat and the leaps of the susceptibility are more obvious, and the transition points approach the exact critical point.}
    \label{fig:2-3-d-Cv-Chi}
\end{figure*}

In this section, we apply the previous variational ansatz design procedure to the long-range interacting Ising model, where we prepare CPS $\ket{\tilde{+}}$ as the initial state. Equipped with the thermal state, we can calculate the specific heat $C_v$ and susceptibility $\chi$ as a function of $K\equiv J\beta$. The measurement process for $C_v$ and $\chi$ is shown schematically in Figure~\ref{fig:measurement-process}. The QITE-ansatz algorithm for Ising model is compactly shown in Algorithm~{\ref{alg:qite-ansatz}}. Our numerical simulations are carried out on the Qiskit noiseless statevector quantum simulator~\cite{Qiskit}.

The initial state and variational ansatz are chosen as described in section~\ref{sec:QITE}. To calculate the thermal expectation values of the Ising model, we only need to calculate the imaginary time evolution of the product state $\ket{\tilde{+}}$. With the initial state, and for every local interaction term in Ising model $Z_iZ_j(\forall i,j\in\Lambda$), we have the corresponding relevant Pauli strings
\begin{align}
    Z_iY_j, Y_i Z_j.
    \label{eq:ZZ_unitary}
\end{align}
Then we can construct the variational ansatz for the target Ising Hamiltonian, i.e., for finite $\alpha$, since the lattice sites are all-to-all coupled with the $Z_iZ_j$ term, the Pauli exponentials gates $e^{-i\theta Z_iY_j}$ and $e^{-i\theta Y_iZ_j}$ are added with all-to-all connection. While for infinite $\alpha$, only nearest-neighbor couplings are considered in the Hamiltonian. So the gates are added only between pairs of nearest neighbor sites.
An example of a variational ansatz for nearest-neighbor Ising chain under periodic boundary conditions is shown in figure~\ref{fig:circuit}\textbf{c}. Each layer of the variational ansatz consists of one layer of ZY-Pauli exponentials and one layer of YZ-Pauli exponentials, as shown in the dashed box. The Pauli exponentials are ladder-arranged, to increase the correlation between sites that can be generated by the variational ansatz. In the figure, we show the case of layers $L=1$. In the following numerical simulations, we use $L=2$ if not specified otherwise. 

We assume the imaginary time evolution of each local interaction term $e^{\tau Z_iZ_j}$ can be realized with the Pauli exponentials $e^{-i\theta Z_i Y_j}e^{-i\theta' Y_i Z_j}$, which have the same support of $Z_iZ_j$. These two Pauli exponentials are enough in the 2-qubit case as indicated by Eq.~(\ref{eq:two-qubit-QITE}), but are not when the system size is large and when the system approaches the critical point, as explained in the previous section. It means that the expressivity of this variational ansatz is not sufficiently good to carry out the whole imaginary time evolution $e^{-\tau H}$. Limited expressivity leads to systematic errors, which will affect the numerical results.

First, we present the numerical results of the nearest-neighbor Ising model(NNIM), i.e., taking the limit $\alpha\rightarrow\infty$ in Eq.~(\ref{eq:long-range-interacting-Hamiltonian}). With the nearest-neighbor interaction, there are $N=2D|\Lambda|L$ parameters in the variational ansatz. In two and three-dimensional NNIMs, there is a second-order phase transition in the infinite volume limit, where the critical points are $K_c=\ln(1+\sqrt{2})/2\approx 0.441$~\cite{Schultz_64} and $0.222$~\cite{Sonsin_2015} for dimension $D=2,3$, respectively. The specific heat and susceptibility would hence diverge near the critical point in the infinite volume limit. Figure~\ref{fig:2-3-d-Cv-Chi} shows the specific heat and susceptibility for various $K$ values obtained via QITE-ansatz. The lattice size is $2\times 2,3\times 3,4\times 4$ for the 2-D system, marked by triangular-down, circle and triangular-up, respectively, and $2\times 2\times 2,3\times 3\times 2$ for the 3-D system, with results marked by triangular-down and circle respectively. In the evolution of the variational parameters, we use the Euler method with step length $\delta\tau=0.002$ as in Eq.~(\ref{eq:Euler-integration}), which is chosen such that further shrinking the step length has no impact on the numerical results (We will take this step length also throughout the following simulations.). We see that the QITE results converge well with the results from exact diagonalization(ED) when the system size is small for both 2-D and 3-D systems. For $4\times 4$ and $3\times 3\times 2$ lattices, the specific heat curves deviate from the ED curves near the critical point, which result from the limitation of the variational ansatz expressivity. The expressivity can be improved by increasing the number of ansatz layers and using longer Pauli strings for each local interaction term beyond $Z_iY_j, Y_i Z_j$. More detailed error analyses are shown in Appendix~\ref{app:error-and-layer-estimation}.

Indications of the Ising criticality can be observed in figure~\ref{fig:2-3-d-Cv-Chi}. The critical temperatures of 2-D and 3-D systems in the infinite volume limit are denoted by the black dashed line. Near the critical points, values of the specific heat and susceptibility increase, and there are peaks in the specific heat as a function of $K$. For 2-D NNIM with volume $N_d\times N_d$, we denote the position of the peak as $K_c(N_d)$. For larger system sizes, $K_c(N_d)$ moves slowly towards the infinite volume critical point $K_c$. To guide the eye of this movement, we draw the grey solid line in the upper left panel of figure~\ref{fig:2-3-d-Cv-Chi}. The analytic expression of the grey solid line is inspired by the finite size scaling~\cite{Pascal_lecture}.

\begin{figure}
    \centering
    \includegraphics[width = 0.45\textwidth]{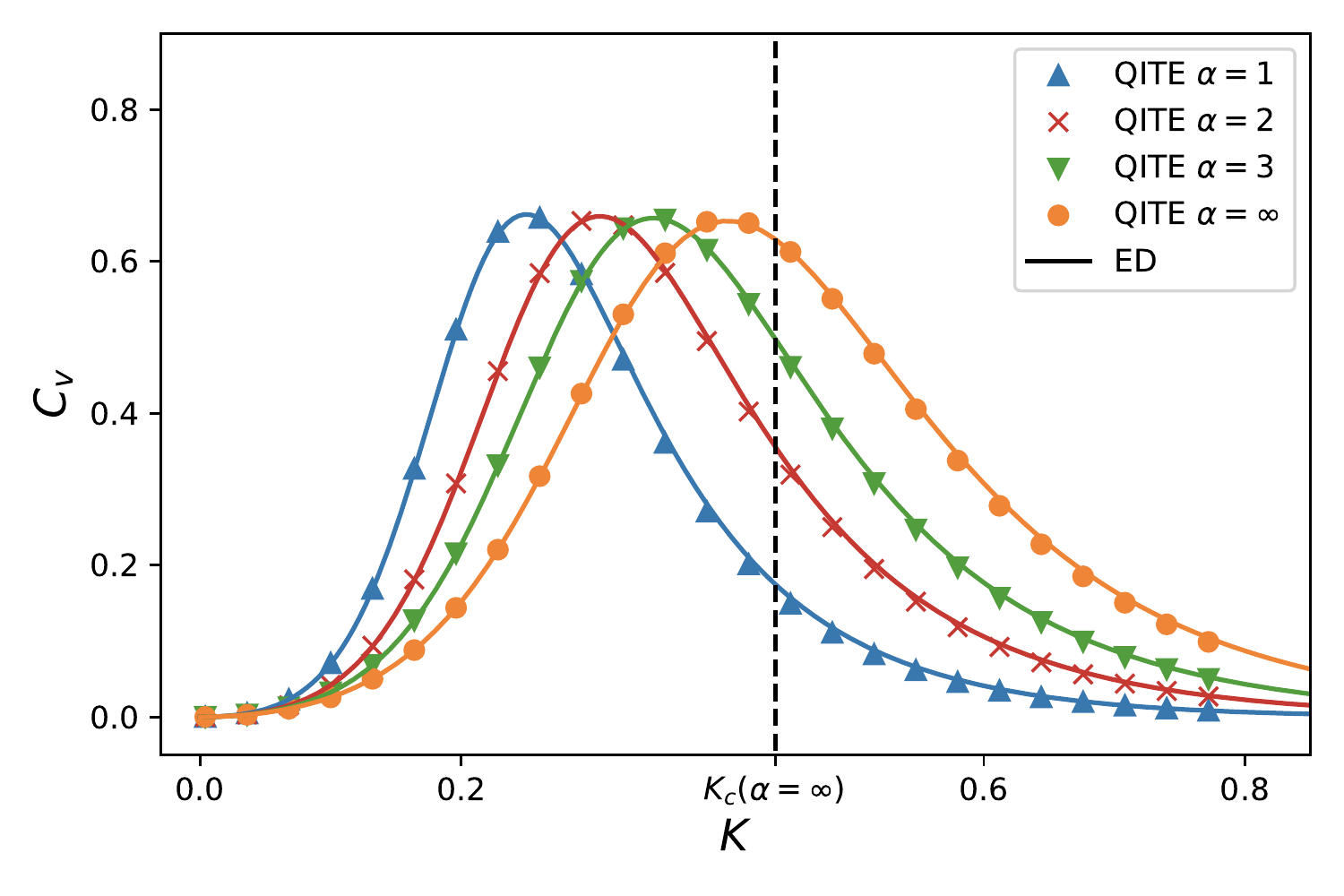}
    \caption{Specific heat as a function of $K$ in the 2-D long-range interacting Ising model with interaction range $\alpha=1,2,3,\infty$, where smaller $\alpha$ indicates larger interaction range. The system size is $|\Lambda|=3\times 3$. ED represents results from exact diagonalization. We see that for various $\alpha$ and $K$, the QITE results and the ED results are consistent. The black dashed line denotes the exact critical point of the 2-D NNIM in the infinite volume limit. As $\alpha$ decreases, the peak of the specific heat curve left shift, indicating that the effective dimension is raised for a larger interaction range.}
    \label{fig:specific-heat-long-range}
\end{figure}

Figure~\ref{fig:specific-heat-long-range} presents the behavior of the specific heat for the 2-D long-range interacting Ising model with finite $\alpha$. Compared with the nearest neighbor interaction, the long-range model introduces more $Z_iZ_j$ interactions and requires more variational parameters. There are $N=|\Lambda|(|\Lambda|-1)L$ parameters in the variational ansatz. The system size in the figure is $|\Lambda|=3\times 3$, with $\alpha=1,2,3$ and the nearest neighbor case $\alpha=\infty$, marked by the triangular-up, cross, triangular-down and circle, respectively. We see that for various $\alpha$ and $K$, the QITE-ansatz results and the ED results are consistent. Moreover, the peak of the specific heat shifts to the direction of high temperature(smaller $K$) for a larger interaction range(smaller $\alpha$). This behavior is reasonable since the long-range interaction effectively raises the system's dimension, and a higher system dimension leads to a higher critical temperature, e.g., 3-D NNIM critical temperature is higher than that of 2-D NNIM.

\section{Discussion}\label{sec:discussion-and-outlook}
This work discuss the possibility of using the imaginary time evolution algorithm to prepare the thermal state of the Ising model on NISQ devices. We numerically calculate the specific heat and susceptibility of the long-range interacting Ising model with the prepared thermal state. We find that the results using the quantum algorithm are consistent with the ones from exact diagonalization for various temperatures, including the critical and low-temperature regions.

We present a systematic procedure to design a variational ansatz for the thermal state preparation. This ansatz is inherited from the quantum circuits used in QITE-measure algorithm. We show that it out-performs the original circuit designed using QITE-measure. This variational ansatz can be further simplified according to the symmetries of the Hamiltonian and the initial state.

In our numerical simulation results, indication of critical behavior can be observed in the calculation of heat capacity and susceptibility of 2-D and 3-D Ising model. The universality properties of Ising model including the critical exponents can be extracted from these quantities in the thermal dynamic limit, where larger Ising system should be simulated to approach the limit. Larger Ising system simulations resort to more advanced quantum devices with more qubits and less error, which are hopefully to be experimentally realized in the near future.

The ideas proposed in this work can be applied to study the critical behavior of other classical models, such as the $Q$-state Potts model, which would be difficult to simulate using the Monte-Carlo algorithm when $Q$ is very large. Additionally, according to the correspondence of the $D$ dimensional quantum model to the $D+1$ dimensional classical model~\cite{Hsieh2012FromDQ}, the algorithm can also be used to study quantum phase transition.

\section{Acknowledgements}
We thank Xiao Yuan, Jinzhao Sun, Lena Funcke, Stefan K\"uhn and Yahui Chai for helpful discussions. X.W. and X.F. were supported in part by NSFC of China under Grants No. 12125501, No. 12070131001, and No. 12141501, and National Key Research and Development Program of China under No. 2020YFA0406400. PS acknowledges support from: ERC AdG NOQIA; Ministerio de Ciencia y Innovation Agencia Estatal de Investigaciones (PGC2018-097027-B-I00/10.13039/501100011033,  CEX2019-000910-S/10.13039/501100011033, Plan National FIDEUA PID2019-106901GB-I00, FPI, QUANTERA MAQS PCI2019-111828-2, QUANTERA DYNAMITE PCI2022-132919,  Proyectos de I+D+I “Retos Colaboración” QUSPIN RTC2019-007196-7); MICIIN with funding from European Union NextGenerationEU(PRTR-C17.I1) and by Generalitat de Catalunya;  Fundació Cellex; Fundació Mir-Puig; Generalitat de Catalunya (European Social Fund FEDER and CERCA program, AGAUR Grant No. 2021 SGR 01452, QuantumCAT \ U16-011424, co-funded by ERDF Operational Program of Catalonia 2014-2020); Barcelona Supercomputing Center MareNostrum (FI-2022-1-0042); EU Horizon 2020 FET-OPEN OPTOlogic (Grant No 899794); EU Horizon Europe Program (Grant Agreement 101080086 — NeQST), National Science Centre, Poland (Symfonia Grant No. 2016/20/W/ST4/00314); ICFO Internal “QuantumGaudi” project; European Union’s Horizon 2020 research and innovation program under the Marie-Skłodowska-Curie grant agreement No 101029393 (STREDCH) and No 847648  (“La Caixa” Junior Leaders fellowships ID100010434: LCF/BQ/PI19/11690013, LCF/BQ/PI20/11760031,  LCF/BQ/PR20/11770012, LCF/BQ/PR21/11840013). Views and opinions expressed in this work are, however, those of the author(s) only and do not necessarily reflect those of the European Union, European Climate, Infrastructure and Environment Executive Agency (CINEA), nor any other granting authority.  Neither the European Union nor any granting authority can be held responsible for them.

\appendix

\section{Simplification of thermal state preparation in classical field theory}\label{app:cps-expectation}

The Hamiltonian of a classical field theory is naturally diagonalized and can be written as a linear combination of Pauli-$Z$ operators, such as the Ising model considered in the main text and $Q$-state Potts model. Such Hamiltonian has energy eigenstates that can be encoded on the computational basis of qubits, and all the Pauli-$Z$ operators commute with each other. To compute the expectation values of such Hamiltonian's thermal state,  we only need imaginary time evolution on an initial state $\ket{\tilde{+}}\equiv\ket{+}^{\otimes N_q}$ where $N_q$ is the number of system's qubits and $\ket{+}=(\ket{0}+\ket{1})/\sqrt{2}$. A similar idea is also proposed in the tensor network algorithm targeting on classical Ising model~\cite{Verstraete_06}. The above statement is proved as follows.

The thermal expectation values $\langle O\rangle$ as defined in Eq.~(\ref{eq:thermal-expectation}) can be expanded with an arbitrary orthogonal basis $\{\ket{i}\}$
\begin{align}
    \langle O\rangle =\frac{\sum_i \bra{i}e^{-\tau H}Oe^{-\tau H}\ket{i}}{Z_{2\tau}},
    \label{eq:appendix-thermal-expectation-value}
\end{align}
where 
\begin{align}
    Z_{2\tau} =\sum_i \bra{i}e^{-2\tau H}\ket{i}.
\end{align}
We choose the orthogonal basis of Pauli-$X$ operators $\{\ket{i}\}=\{\ket{+},\ket{-}\}^{\otimes N_q}$. Notice that all vectors in the set can be generated by applying Pauli-$Z$ operators on one basis vector $\ket{\tilde{+}}$. For example
\begin{align}
    Z_2 Z_1\ket{+}_2\ket{+}_1\ket{+}_0=\ket{-}_2\ket{-}_1\ket{+}_0.
\end{align}
The Hamiltonian consists of Pauli-$Z$ operators, so it commutes with all the Pauli-$Z$ operators. Thus, all terms in the partition function are equal
\begin{align}
    \bra{i}e^{-2\tau H}\ket{i}=\bra{\tilde{+}}e^{-2\tau H}\ket{\tilde{+}},
    \label{eq:identical-denominator}
\end{align}
for all $\ket{i}\in \{\ket{+},\ket{-}\}^{\otimes N_q}$, and we have $Z_{2\tau}=2^{N_q}\bra{\tilde{+}}e^{-2\tau H}\ket{\tilde{+}}$. Further, notice that all the observables concerning specific heat and susceptibility in Eq.~(\ref{eq:concerned-observable}) consist of Pauli-$Z$ operators, which can be formally written as 
\begin{align}
    O=\sum_m h_m\tilde{Z}_m,
    \label{eq:observable-Z-decomposition-app}
\end{align}
where $\tilde{Z}_m$ denotes the tensor product of $Z$ operators at some sites and identity operators at others. Similar to Eq.~(\ref{eq:identical-denominator}), all terms in the numerator of Eq.~(\ref{eq:appendix-thermal-expectation-value}) are equal
\begin{align}
    \bra{i}e^{-\tau H}\tilde{Z}_me^{-\tau H}\ket{i}=\bra{\tilde{+}}e^{-\tau H}\tilde{Z}_me^{-\tau H}\ket{\tilde{+}},
    \label{eq:identical-numerator}
\end{align}
for all $\ket{i}\in \{\ket{+},\ket{-}\}^{\otimes N_q}$. Thus we have 
\begin{equation}
\begin{aligned}
     \langle \tilde{Z}_m\rangle&=\frac{2^{N_q}\bra{\tilde{+}}e^{-\tau H}\tilde{Z}_me^{-\tau H}\ket{\tilde{+}}}{Z_{2\tau}}\\
     &=\frac{\bra{\tilde{+}}e^{-\tau H}\tilde{Z}_me^{-\tau H}\ket{\tilde{+}}}{\bra{\tilde{+}}e^{-2\tau H}\ket{\tilde{+}}}.
\end{aligned}
\end{equation}

In conclusion, the thermal expectation value of an observable $O=\sum_m h_m \tilde{Z}_m$ with a thermal state of a classical Hamiltonian can be derived according to imaginary time evolution on initial state $\ket{\tilde{+}}$,
\begin{equation}
\begin{aligned}
     \langle O\rangle&=\sum_m h_m \langle \tilde{Z}_m\rangle\\
     &=\sum_m h_m\frac{\bra{\tilde{+}}e^{-\tau H}\tilde{Z}_me^{-\tau H}\ket{\tilde{+}}}{\bra{\tilde{+}}e^{-2\tau H}\ket{\tilde{+}}}\\
     &=\sum_m h_m \bra{\tilde{+}(\tau)}\tilde{Z}_m\ket{\tilde{+}(\tau)},
\end{aligned}
\end{equation}
where $\ket{\tilde{+}(\tau)}$ is imaginary time evolved state according to Eq.~(\ref{eq:psi_QITE}). The state is initialized as $\ket{\tilde{+}(0)}=\ket{\tilde{+}}$. Thus we prove the statement in Eq.~(\ref{eq:simplified-estimation}).

\section{Error analysis and circuit layers estimation}\label{app:error-and-layer-estimation}

There are four main sources of errors when implementing the QITE-ansatz algorithm on real quantum devices~\cite{LiYing2017}
\begin{itemize}
    \item The variational ansatz has limited expressivity. The imaginary time evolution proceeds on the manifold expanded by the variational ansatz. Thus the evolved wave function deviates from the true wave function in Eq.~(\ref{eq:psi_QITE}), and leads to the systematic error of the expectation values of the observables.
    \item Errors arise from the numerical integration using the Euler method as in Eq.~(\ref{eq:Euler-integration}).
    \item Noisy quantum gates and readout processes in quantum devices result in systematic errors when evaluating expectation values and estimating $M$ and $V$ (See Eq.~(\ref{eq:M-V-calculation})).
    \item Finite number of shots results in statistical errors in evaluating expectation values, $M$ and $V$.
\end{itemize}
Errors from the first and the second items are specific to the QITE-ansatz algorithm. The third and forth errors exist in general for any quantum algorithms. In the following contents, we will discuss these errors in detail.

\begin{figure}
    \centering
    \includegraphics[width = 0.45\textwidth]{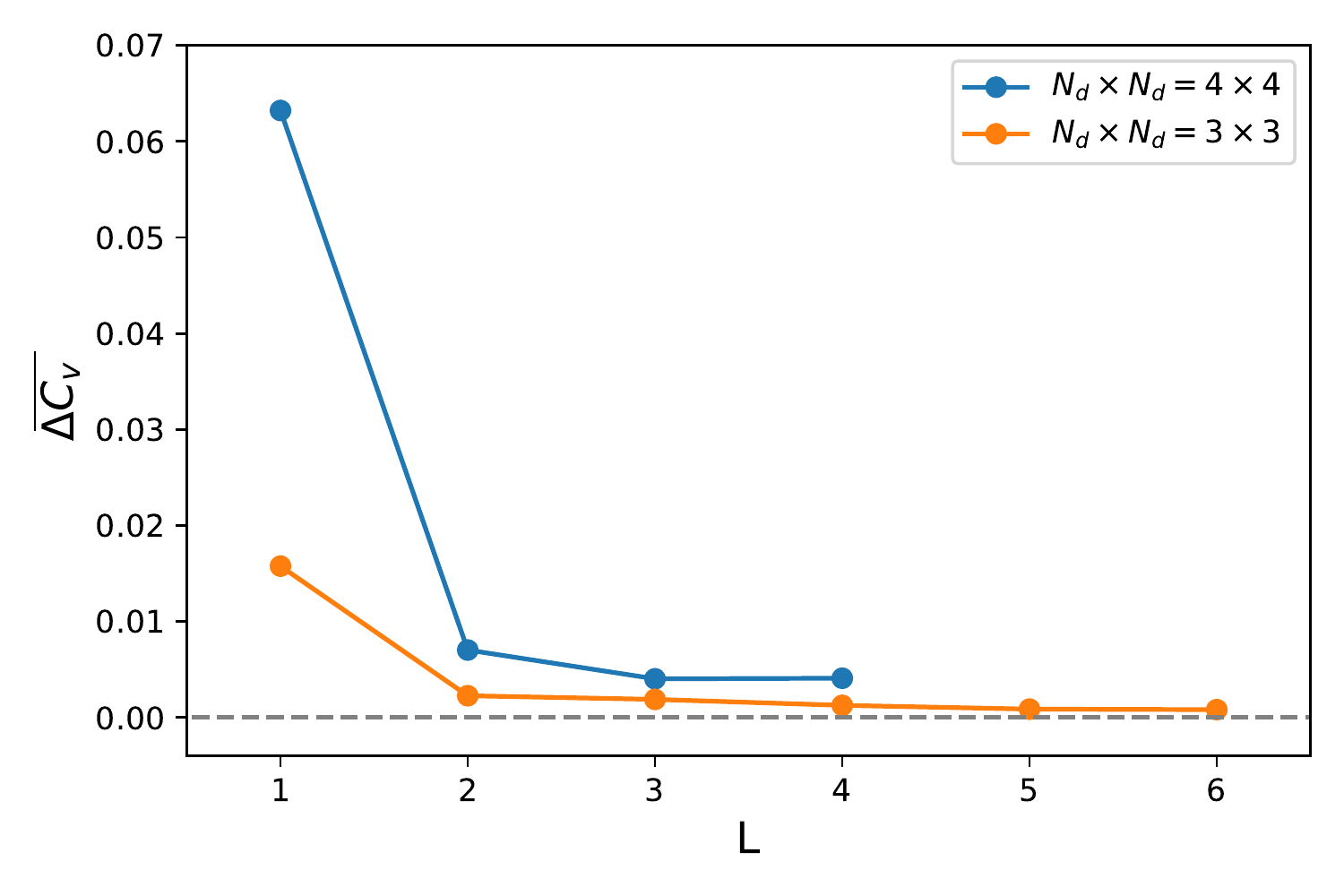}
    \caption{The average absolute error of specific heat as a function of variational ansatz layers. We utilize the 2-D nearest neighbor Ising model with two volumes $|\Lambda|=N_d\times N_d=3\times 3, 4\times 4$. The limitation of variational ansatz can be well controlled by increasing the layers. As the number of layers is larger than some transition layers $L^*$, the error has no obvious change. Theoretically, we can predict $L^*\in[1.5,3]$ for $N_d=3$ and $L^*\in[2,4]$ for $N_d=4$.}
    \label{fig:error-layer-dependence}
\end{figure}

\begin{figure*}
    \centering
    \includegraphics[width = 0.8\textwidth]{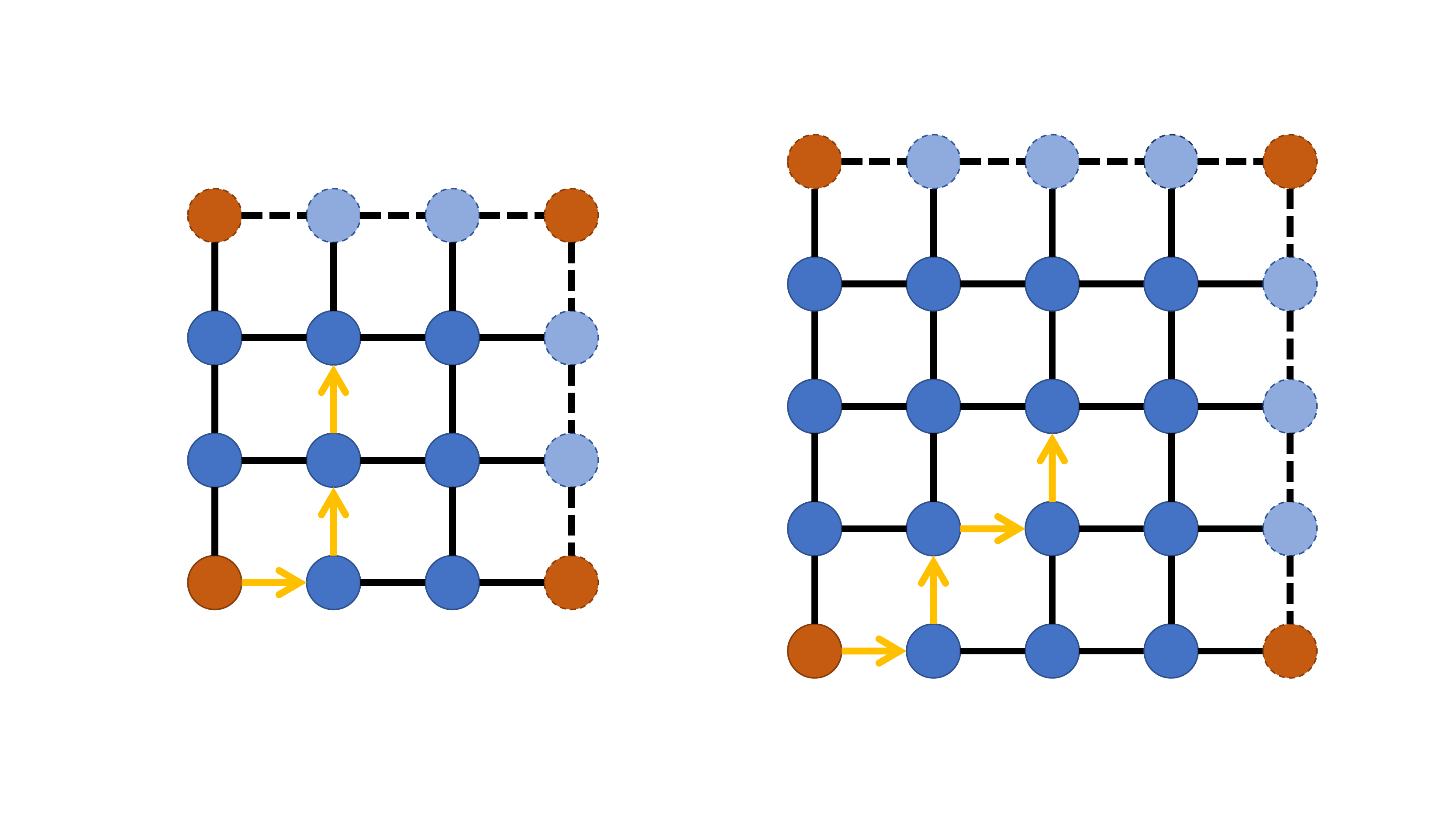}
    \caption{An Illustration to explain the transition layer in figure~\ref{fig:error-layer-dependence}. We plot the 2-D nearest neighbor Ising model with two volumes $|\Lambda|=3\times 3, 4\times 4$. Circles with solid edges are real spins. Circles with dashed edges are virtual spins from the periodic boundary condition. The spin at the origin(lower left corner) is denoted by the orange circle, which has three corresponding dashed-edge circles at three other corners. The yellow arrows denote the shortest path connecting the most remote spin to the spin at the origin. One layer of the variational ansatz generates the correlation between two spins at least 1 unit length apart. With at most $DN_d/2$ layers, the variational ansatz can connect all the spins in the graph.}
    \label{fig:L_transition}
\end{figure*}

The errors from the limited variational ansatz expressivity have been shortly discussed in the main text. There are two ways to improve expressivity. The first is by increasing the number of ansatz layers, and the second is by considering longer Pauli strings expansion in Eq.~(\ref{eq:Hermitian-expansion}) for each local interaction term in the Hamiltonian. It is not hard to see that by extending number of layers to infinity and taking the expansion on the whole system, the variational ansatz can carry out the evolution $e^{-\tau H}$ exactly. In the following text, we numerically investigate how these two aspects affect the performance in calculating the specific heat of 2-D NNIM.

\begin{figure}
    \centering
    \includegraphics[width = 0.45\textwidth]{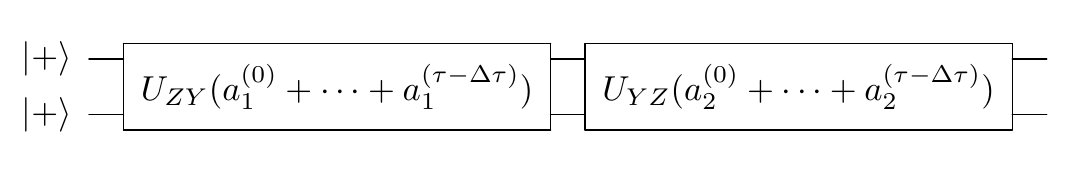}
    \caption{The quantum circuit equivalent to the QITE-measure circuit in figure~\ref{fig:circuit}\textbf{a}. Here $U_{ZY}(\theta)\equiv e^{-i\theta ZY},U_{YZ}(\theta)\equiv e^{-i\theta YZ}$. This circuit can be expressed perfectly with only one layer of the variational ansatz in figure~\ref{fig:circuit}\textbf{b}.}
    \label{fig:circuit-appendix}
\end{figure}

The limitation of finite ansatz layers can be observed by tuning the number of layers $L$. In figure~\ref{fig:error-layer-dependence}, we compute the average absolute error of 2-D NNIM specific heat as a function of $L$, in case of lattice volumes $|\Lambda|=3\times 3, 4\times 4$. The average absolute error is defined by
\begin{align}
    \overline{\Delta C_v}\equiv \frac{1}{|K_{\max}-K_{\min}|}\int_{K_{\min}}^{K_{\max}} \ud K~|C_v-C_v^{ED}|,
\end{align}
where $C_v$ is specific heat from the quantum simulator, and $C_v^{ED}$ is from exact diagonalization. Here we take the integration range $[K_{\min},K_{\max}]=[0,1]$. The errors of specific heat decrease rapidly as $L$ increase and saturate to a platform after a certain layer $L^*$. We will analyze this transition layers $L^*$ after a while. When $L>L^*$, the remaining average absolute error of specific heat is mainly from the finite length of Pauli strings expansion.

Here provide an empirical explanation of the transition layer $L^*$, as observed in figure~\ref{fig:error-layer-dependence}. It also helps to estimate how many layers we need when constructing variational ansatz for simulating NNIM. As shown in figure~\ref{fig:L_transition}, variational ansatz generates correlation in the spin system. In the best case, the correlation between two neighboring spins is generated by one unitary transformation such as $e^{-i\theta ZY}$ in the Ising case; In the worst case, we need a whole layer of the variational ansatz such as $e^{-i\theta ZY}e^{-i\theta' YZ}$ to generate such correlation. The transition layer $L^*$ indicates the lowest number of circuit layers to generate correlation between the two most distant spins in the D-dimensional nearest neighbor lattice system. Thus for D-dimensional NNIM with volume $N_d^D$ and PBC, as the Manhatten distance of the most remote two spins is $DN_d/2$ (Equal to the number of the yellow arrows in figure~\ref{fig:L_transition}, where $D=2, N_d=3,4$ respectively.), the transition layer would be in the range
\begin{align}
    \frac{DN_d}{2G}\leq L^* \leq \frac{DN_d}{2},
    \label{eq:layers-required}
\end{align}
which corresponds to the best case and worst case mentioned above. Here $G$ is the number of Pauli exponentials in one layer, i.e., the number of relevant Pauli operators for some local interaction terms. The transition layers in figure~\ref{fig:error-layer-dependence} are in accord with this range, i.e., $N_d/2\leq L^* \leq N_d$, and we see larger number of layers have almost no improvement to the average absolute error of the specific heat. Thus we say $L^*$ layers are enough for variational ansatz to simulate NNIM. This estimation on the number of ansatz layers can be generalized to more complicated short-range interacting models.

Comparing the required number of layers of the variational ansatz provided by Eq.~(\ref{eq:layers-required}) and the layers of quantum circuits used in QITE-measure, one finds the former is much less than the latter. It can be partially explained using the example of the two-qubit Ising system shown in the main text. For the QITE-measure circuit figure~\ref{fig:circuit}\textbf{a}, due to the commutability of relevant Pauli operators $[ZY, YZ]=0$, it is equivalent to the circuit shown in figure~\ref{fig:circuit-appendix}, which consists of only two Pauli exponentials where the rotation angles are the summation of all the coefficients of the corresponding Pauli exponentials in figure~\ref{fig:circuit}\textbf{a}. Therefore, if we use one layer of the circuit in figure~\ref{fig:circuit}\textbf{b}, and $\theta_1=a_1^{(0)}+\ldots+a_1^{(\tau-\Delta \tau)},\theta_1'=a_2^{(0)}+\ldots+a_2^{(\tau-\Delta \tau)}$, the QITE-measure circuit could be rephrased without loss of the precision. Thus compared with the QITE-measure circuit, the number of variational ansatz layers used in our simulation can be significantly reduced.

Numerical integration errors can be controlled via a more elaborate numerical integration algorithm. In the main text, we use the Euler method that accumulates a global error of $\mathcal{O}(\delta \tau)$ at the final step. One could use a more elaborate numerical algorithm such as the 4th-order Runge-Kutta method to control the systematic error, which accumulates a global error of $\mathcal{O}(\delta \tau^4)$ at the final step. In our simulations, as the numerical integration error is not the dominate systematic error, the Euler method is sufficiently good. 

Errors from noisy quantum gates and readout processes lead to systematic deviations of the measurement results to the noiseless ones.  For NISQ devices, there are many error mitigation techniques to reduce these deviations. For example, errors from two-qubit gates can be mitigated by zero-noise extrapolation~\cite{Temme_2017,LiYing2017} and quasi-probability decomposition~\cite{LiYing2017,Suguru2018}. The readout error can be mitigated by classical bit-flip correction~\cite{Lena2022,Berg2022}. The error mitigation techniques reduce the systematic deviations of the noisy results to the noiseless ones and shed light on the real applications of NISQ devices~\cite{Kim2023,Cao2023}.

Finite number of shots error is a statistical error, which can be suppressed by increasing the number of shots. In the measurement procedure, the observables are split into a weighted sum of Pauli operators and each can be measured separately, at the cost of many shots. The number of shots can be reduced by collecting mutually commuting Pauli operators together before measuring all operators within a collection simultaneously~\cite{Kandala2017, Crawford2021efficientquantum}. In the measurement process of the Ising model in Figure~\ref{fig:measurement-process}, the weighted Pauli operators are mutually commuting Pauli-$Z$ strings. They can be measured simultaneously in computational basis.

\section{Execution time of the nearest-neighbor Ising model}

In this appendix, we estimate the execution time to study the critical behavior of the nearest-neighbor Ising model (NNIM) using the QITE-ansatz algorithm. 

The execution time of the QITE-ansatz algorithm can be estimated by the number of steps of the imaginary time evolution, times the number of expectation values evaluated per step, times the number of two-qubit quantum gates (Assume no parallelization of the quantum gates, and the two-qubit gates dominate the execution time), i.e., 
\begin{align}
    \mathrm{time}\sim (\mathrm{steps})\times (\# \mathrm{~of~expectations})\times (\# \mathrm{~of~gates}).
\end{align}
Among the three factors, if the critical temperature of the system is of $\mathcal{O}(1)$, then the total evolution should have the same order, and the number of evolution steps is a constant, which is the case of simulating NNIM.

For each step of the time evolution, one needs to estimate the expectation values of entries of the $M$ matrix and $V$ vector. Thus 
\begin{align}
    \# \mathrm{~of~expectations}\sim \mathcal{O}(N^2),
    \label{eq:num-of-expectations}
\end{align}
 where $N$ is the dimension of the $M$ matrix, equals the number of variational parameters. For the variational ansatz proposed in this work, the number of variational parameters is $N=2D|\Lambda|L$ for NNIM, where $D$ is the spatial dimension of the NNIM lattice $\Lambda$, $|\Lambda|$ is the lattice size, $L$ is the number of ansatz layers. In Appendix~B of the manuscript, we estimate the number of layers required for NNIM, where $L\sim \mathcal{O}(DN_d)$, $N_d$ is the side length of $\Lambda$. The lattice size and the side length are related by $|\Lambda|=N_d^D$. Thus, the number of expectation evaluations reads
\begin{align}
    \# \mathrm{~of~expectations}\sim \mathcal{O}(D^4 N_d^{2D+2}).
\end{align}

To count the number of two-qubit quantum gates, we assume that the CNOT gate is the basic two-qubit gate that can be realized on quantum devices, and the CNOT gate can be applied to every two-qubit pair. Then, the number of CNOT gates in the proposed ansatz is proportional to its number of variational parameters $N$, so that
\begin{align}
\mathrm{\#~of~gates}\sim \mathcal{O}(D^2 N_d^{D+1}).
\end{align}

In summary, the execution time of the QITE-ansatz algorithm for D-dimensional NNIM reads
\begin{align}
    \mathrm{time}\sim\mathcal{O}(D^6 N_d^{3D+3}),
\end{align}
which is a polynomial function of the system size.

There are some improvement methods of the QITE-ansatz algorithm to reduce the execution time, such as the DualQITE algorithm proposed in reference~\cite{gacon2023variational}. Using this algorithm, one can reduce the number of expectation evaluations in Eq.~(\ref{eq:num-of-expectations}) from $\mathcal{O}(N^2)$ to $\mathcal{O}(N)$, and the total execution time is correspondingly reduced to 
\begin{align}
\mathrm{time}\sim\mathcal{O}(D^4 N_d^{2D+2}).
\end{align}

\bibliographystyle{unsrt}
\bibliography{main}
\end{document}